\documentclass[aps,prb,11pt,amsmath,amssymb]{revtex4}

\usepackage{bm}
\usepackage{mathrsfs}                                                                                
\usepackage{graphics}
\usepackage{epsfig}
\usepackage{dcolumn}

\begin{document}
                                                                                                                         
\title{Quantum equations for vibrational dynamics on metal surfaces.}
                                                                                                                           
                                                                                                                           
\vskip0.2cm

\author{Vinod Krishna\footnote{Present Address: Department of Chemistry, 
University of Utah, 315 S 1400 E, UT 84112-0850; email:vkrishna@hec.utah.edu}}
\affiliation{Department of Physics, Yale University, New Haven, CT 06520}
\date{\today}
\begin{abstract}   
    A first-principles treatment of the vibrational dynamics of molecular chemisorbates on
metal surfaces is presented. It is shown that the mean field quantum evolution of the 
vibrational position operator is determined by a quantum Langevin equation with an electronic 
friction. In the mean field limit, the quantum noise and friction are related by the quantum 
fluctuation-dissipation theorem. The classical limit of this model is shown to agree with 
previously proposed models. A criterion is presented to describe the validity of the
weak-coupling approximation and equations of motion for the dynamics in the presence of 
strong nonadiabatic coupling to electron-hole pairs are presented.
\end{abstract}
\maketitle
 


\baselineskip 13pt
\baselineskip 16pt
 
 

 
 

\section*{I. INTRODUCTION}
      Understanding the nonadiabatic dynamics of molecules has been an important and active 
area of research for several decades. Several methods have been developed to study
nonadiabatic molecular dynamics. Among them, mixed quantum-classical methods 
have had considerable success in studying nonadiabatic effects in molecular systems. 
However, fully quantum treatments of nonadiabatic molecular dynamics are few and limited 
in their applicability. A special class of systems, where the nonadiabatic coupling between 
electron and molecular degrees of freedom is important, involve the dynamics of molecules 
chemisorbed on metal surfaces. The motion of molecules in an electron-hole pair bath  
correspond to a limiting case of nonadiabatic dynamics. Molecular degrees of freedom in these 
situations are coupled to an infinite reservoir of electronic excitation modes. In contrast, 
gas phase molecular processes involve nonadiabatic coupling between finite numbers of 
Born-Oppenheimer potential energy surfaces. Due to the continuum nature of the electron 
spectrum near a metal surface, the nonadiabatic coupling of molecular vibrational modes to 
electronic excitations results in the dissipation of vibrational excitations. 
\newline
\par  The dynamics of a chemisorbed molecule on a metal surface is influenced by many different 
processes. At low temperatures, a dominant mechanism for energy transfer between 
the molecule and the surface is the dissipative dynamics induced by electron-hole pair 
excitations. The nonadiabatic coupling between vibrational and electron-hole pair excitations 
determines the relaxation rate of vibrational excitations on the 
surface\cite{Persson}$^{-}$\cite{Tully}. Furthermore, this nonadiabatic coupling also 
influences the dynamical evolution of the vibrational mode on the surface. As a result, 
electron-hole pair excitations play an important role in chemical processes that take 
place on the surface.  
\newline
\par Though there has been considerable theoretical and experimental work devoted to 
understanding the role of electron-hole pair excitations in vibrational relaxation processes 
on metal surfaces\cite{Persson}$^{-}$\cite{guo}, there is no first principles theoretical 
framework which can be used to systematically analyse the role of electron-hole pair excitations
on molecular dynamics. This is true even for the case where electron-hole pair effects on the 
dynamics enter in a mean field fashion. Previously, attempts have been made to guess the formal 
descriptions of the vibrational dynamics of the molecular system in the presence of electron-hole 
pair excitations\cite{Schiach}$^{-}$\cite{gunnar}. Intuitively, because of the continuum nature of 
the electronic spectrum at metal surfaces, the effect of electron-hole pair coupling to vibrational 
dynamics can be thought to be equivalent to an electronic friction acting on the vibrational modes. 
Previous attempts to study the reduced classical dynamics of the vibrational modes have used these 
intuitive ideas to construct Langevin type descriptions of the classical vibrational dynamics. A more 
rigorous approach has been attempted\cite{Tully1} through the application of the Ehrenfest 
method of mixed-quantum classical dynamics. In this approach, effective Langevin type equations 
for the vibrational dynamics of the chemisorbed molecule were derived. However, such mixed 
quantum-classical descriptions are limited to treating the dynamics with the surface at 
zero temperature. Consequently, a fluctuation-dissipation relation needs to be imposed on the 
reduced dynamics of the system. Furthermore, because of the somewhat ad-hoc nature of these 
formulations, there is no quantitative criterion to determine the range of nonadiabatic 
electron-hole pair coupling strengths excitations for which such descriptions are valid. 
As a result, it is desirable to formulate a first principles quantum mechanical formulation of 
the reduced dynamics of molecular chemisorbates on a metal surface. 
\newline
\par This work develops, for the first time, a formal and fully quantum description of the 
nonadiabatic dynamics of molecular modes in the presence of coupling to electron-hole pair 
excitations. It is shown that as long as the number of possible electronic excitations around 
the surface are unlimited, and it is possible to retain a mean field separation between vibrational 
and electronic excitations, the quantum dynamics of the molecule is governed by a generalized
Langevin equation with a quantum noise spectrum. The friction and noise correlation functions 
are shown to be related to each other by the generalized Callen-Welton fluctuation-dissipation 
theorem for quantum systems\cite{Callen,Hanggi}. Finally, it is shown that this 
description of the system dynamics reduces to the mixed-quantum classical Langevin equation 
derived\cite{Tully1} by Head-Gordon and Tully in the limit that the energy of electron-hole 
pair excitations are small compared to the temperature of the surface. This paper also provides a 
quantitative criterion to describe the range of validity of the mean field approximation. 
When the mean field approximation is no longer valid, the resulting nonlinear equations of 
motion describe molecular dynamics in an effective electromagnetic field induced by electron-hole 
pair excitations.
\newline
\par   The paper is outlined as follows. In Sec II. the general Hamiltonian of the system is 
written down in a Born-Oppenheimer basis of wavefunctions, and a transformation is introduced
which converts the electron-hole pairs into a collection of bosonic excitations. The Heisenberg
equations of motion are then written down and the electron-hole pair excitations integrated out.
In Sec.III, a quantum generalized Langevin equation is derived and the fluctuation-dissipation 
relations derived. The mixed quantum classical limit is then discussed. Finally, an approximate 
criterion is presented to estimate the validity of the mean field approximation for a vibrationally
excited molecule adsorbed on the surface. Further generalizations of the equations of motion in the 
regime of strong adiabaticity are then presented. The paper ends with a discussion of the results 
and future work. 

\section*{II. THEORY}
\subsection*{A. The Hamiltonian}
   The Hamiltonian for a molecular vibration on a metal surface can be written as
\begin{equation}
   H = \frac{P^{2}}{2M} + U(R) + H_{e}({\bf{r}}) + V({\bf{r}},{\vec{R}})
\end{equation}
   The first two terms on the right hand side correspond to the Hamiltonian for the vibrational
modes given by the operator $R$. $H_{e}(\bf{r})$ is the electronic Hamiltonian and $V(\bf{r},R)$ 
the coupling between the vibrational and electronic modes. If non-adiabatic effects are weak enough 
to retain some of the separation in timescales between the vibrational and electronic excitations, 
they can be treated as perturbations to the adiabatic motion of the nuclear modes. In such a 
situation a Born-Oppenheimer basis of electron wavefunctions can be defined by
\begin{equation}
  [H_{e} + V({\bf{r}},\vec{R})]\mid\phi_{\bf{k}}\rangle = \epsilon_{k}(\vec{R})\mid\phi_{\bf{k}}\rangle
\end{equation}
 The basis of wavefunctions $\phi_{k}$ depends adiabatically on the vibrational coordinates $R$. 
The vibrational momentum operator is off-diagonal in this basis of wavefunctions. This off-diagonal 
coupling is responsible for non-adiabatic corrections to the Born-Oppenheimer solution of the system.
In this basis, the total Hamiltonian can be rewritten as 
\begin{equation}
H = \frac{P^{2}}{2M} + U(R) + \sum_{k}{\epsilon_{k}(R)\mid\phi_{k}\rangle\langle\phi_{k}\mid} 
 + \sum_{k,l}{\mid\phi_{k}\rangle\langle\phi_{k}\mid\frac{P^{2}}{2M}\mid\phi_{l}\rangle
\langle\phi_{l}\mid} 
\end{equation}
 The off-diagonal coupling of the vibrational kinetic energy can be simplified as 
\begin{equation}
 \langle\phi_{k}\mid\frac{P^{2}}{2M}\mid\phi_{l}\rangle = -\frac{i\hbar}{M}\langle\phi_{k}\mid
\nabla_{R}\mid\phi_{l}\rangle\cdot\vec{P} -\frac{\hbar^{2}}{2M}\langle\phi_{k}\mid\nabla_{R}^{2}\mid
\phi_{l}\rangle
\end{equation}
  The term which consists of the second derivative of the basis states is neglected and the 
non-adiabatic coupling simplifies to a term linear in the vibrational momentum. With this 
approximation the Hamiltonian of the system becomes
\begin{equation}
H = \frac{P^{2}}{2M} + U(R) + \sum_{l}{\epsilon_{l}(R)\mid\phi_{l}\rangle\langle\phi_{l}\mid}
     + \sum_{k,l}{\mid\phi_{k}\rangle \vec{V}_{kl}(R)\langle\phi_{l}\mid}\cdot
\frac{\vec{P}}{M}  
\end{equation}
   The off-diagonal coupling $\vec{V}_{kl}$ is a matrix whose diagonal elements are zero. It is 
defined as 
\begin{equation}
  \vec{V}_{kl}(R) = -i\hbar\langle\phi_{k}\mid\nabla_{R}\phi_{l}\rangle
\end{equation}
  It is also convenient to define 
\begin{equation}
   \vec{D}_{kl}(R) = \langle\phi_{k}\mid\nabla_{R}\phi_{l}\rangle
\end{equation}
    The off-diagonal coupling is a vector operator which couples to the momentum operator. It is
equivalent to that of a vector potential acting on the system. Thus, the nonadiabatic effects of 
the electronic modes appear as magnetic potentials acting on the molecular degrees of freedom. 
\par
    The rest of the derivation can be sketched out as follows. The Hamiltonian is rewritten 
in terms of electron-hole pair creation and annihilation operators. The Heisenberg equations 
of motion are then derived for the vibrational coordinate and the electron-hole pair operators.
They are then solved self-consistently for the electron-hole pair operators, and the electronic 
degrees of freedom are eliminated in a procedure reminiscent of the methods used in Ref. (33), to 
describe the vibrational dynamics. The equations of motion so obtained are shown to correspond 
to a generalized quantum Langevin equation\cite{Gardiner,Zoller}. 
\subsection*{B. The Boson Approximation}
\par
    The Hamiltonian in Eq.(3) can be rewritten in a second quantized form by introducing 
creation and annihilation operators to describe electronic states. For a molecule adsorbed on 
a metal surface, only electronic excitations corresponding to excitations from filled orbitals 
below the Fermi shell to unoccupied orbitals above the Fermi shell contributes. This implies 
that only terms with $k > k_{f} > l$ contribute to the summation over off-diagonal terms. Thus 
the effective Hamiltonian for molecular vibrations coupled to electron-hole pair excitations is 
\begin{equation}
 H = \frac{P^{2}}{2M} + U(R) + \sum_{k}{\epsilon_{k}(R)c_{k}^{\dagger}c_{k}} + 
\sum_{k>k_{f}>l}{[\vec{V}_{kl}(R)c_{k}^{\dagger}c_{l} + \vec{V}_{lk}(R)c_{l}^{\dagger}
c_{k}]\cdot\frac{\vec{P}}{M}}
\end{equation}
\par 
    For a metallic system, the number of electron-hole pair excitations is much smaller than the 
number of bulk electrons in the ground state. This is because electron-hole pair excitations depend 
on the number of electrons close to the fermi surface, which is of an order $O(1/N)$ in comparison 
to the total number of electrons $N$ enclosed by the fermi surface. Consequently, one can ignore the 
contributions from processes which do not correspond to simple electron-hole pair excitations. Since 
the energy spectrum of the metal surface near the fermi level is continuous, the number of possible 
electron-hole pair excitations is for all practical purposes, unlimited. Consequently, fermion pair 
operators of the form $c_{k}^{\dagger}c_{l}$ with $k>l$ can be regarded as having an approximately 
bosonic character\cite{tom,Ram}. In other words, the pair operators are replaced by
\begin{equation}
   c_{k}^{\dagger}c_{l} = \rho_{kl}^{\dagger}
\end{equation}

    Since the index $k$ is always greater than the index $l$, the pair operators obey the 
commutation relations, 
\begin{equation}
  [c_{l}^{\dagger}c_{k}, c_{n}^{\dagger}c_{m}] = 0
\end{equation}
   
    Here $c_{n}^{\dagger}c_{m}$ is another electron-hole pair annihilation operator with $n<m$.
 Also, the pair operators approximately satisfy
\begin{equation}
  [c_{l}^{\dagger}c_{k},c_{m}^{\dagger}c_{n}] = \delta_{ln}\delta_{km}
\end{equation} 
 In terms of the operators $\rho_{mn}$ the approximate commutation relations become
\begin{equation}
  [\rho_{kl} ,\rho_{mn}^{\dagger}] = \delta_{mk}\delta_{nl}
\end{equation}
and
\begin{equation}
  [ \rho_{kl},\rho_{mn}] = 0
\end{equation}
     
  Thus, electron-hole pairs created near the Fermi energy can be expected to behave approximately 
as Bosons. By approximating the Fermionic pair operators as Bosonic creation and annihilation operators,
the Hamiltonian for the vibrational motion coupled to electron-hole pair excitations can 
be written as 
\begin{equation}
  H = H_{0}(R)  + \sum_{k>k_{f}>l}{[\vec{V}_{kl}\rho_{kl}^{\dagger} + \vec{V}_{lk}\rho_{kl}]
\cdot\frac{\vec{P}}{M}} + \sum_{k}{\epsilon_{k}(R)c_{k}^{\dagger}c_{k}} .
\end{equation}
   The operators $\rho$ are Bosonic operators which create or destroy electron-hole pair 
excitations. 
\subsection*{C. Equations of Motion} 
   The time dependent Heisenberg operators now satisfy the equations
\begin{eqnarray}
  \dot{\vec{R}}(t) = \frac{i}{\hbar}[H,\vec{R}] \nonumber\\
  \dot{\vec{P}}(t) = \frac{i}{\hbar}[H,\vec{P}] \nonumber\\
\dot{\rho}_{kl}(t) = \frac{i}{\hbar}[H, \rho_{kl}(t)]
\end{eqnarray}
  These equations can be explicitly written out to give
\begin{eqnarray}
M\dot{\vec{R}}(t) = \vec{P}(t) + \sum_{k>k_{f}>l}{[\vec{V}_{kl}\rho_{kl}^{\dagger} + \vec{V}_{lk}\rho_{kl}]} 
\nonumber\\ 
\dot{\vec{P}}(t) = -\nabla_{R}U -\sum_{k}{\nabla_{R}\epsilon_{k} c_{k}^{\dagger}c_{k}} - 
\sum_{k>k_{f}>l}[\nabla_{R}\vec{V}_{kl} \rho^{\dagger}_{kl} + h.c]\cdot\frac{\vec{P}(t)}{M}
\end{eqnarray}
and
\begin{eqnarray}
 \dot{\rho}_{mn}(t) = -i\omega_{mn}\rho_{mn}(t) - \frac{i}{\hbar}\vec{V}_{mn}(t)\cdot\frac{\vec{P}(t)}{M} \nonumber\\
 \dot{\rho}_{mn}^{\dagger}(t) = i\omega_{mn}\rho_{mn}^{\dagger} + \frac{i}{\hbar}\vec{V}_{nm}(t)\cdot\frac{\vec{P}(t)}{M}
\end{eqnarray}
  The energy of the electron excitation from state $n$ below the Fermi surface to state $m$ above
it, has an excitation energy given by
\begin{equation}
   \hbar\omega_{mn}(R) = \epsilon_{m}(R) - \epsilon_{n}(R)
\end{equation} 
  Eq.(16) for the position operator $R(t)$ is differentiated to give
\begin{equation}
M\ddot{\vec{R}}(t) = \dot{\vec{P}} + \sum_{m>n}{[\dot{\vec{V}}_{mn}\rho_{mn}^{\dagger} + \dot{\vec{V}}_{nm}
\rho_{mn}]} +\sum_{m>n}{[\vec{V}_{mn}\dot{\rho}_{mn}^{\dagger} + h.c]}
\end{equation}   
  This can be further simplified by replacing the time derivatives on the right hand side by 
the appropriate commutators with the Hamiltonian. The time derivative of the non-adiabatic coupling
is given by 
\begin{equation}
  \dot{\vec{V}}_{mn}(R) = \frac{i}{\hbar}[H,\vec{V}_{mn}]
\end{equation}  
or
\begin{equation}
  \dot{\vec{V}}_{mn}(R) = \nabla_{R}\vec{V}_{mn}(R)\cdot\frac{\vec{P}(t)}{M} - \frac{ih}{2M}\nabla^{2}_{R}\vec{V}_{mn}
 + \sum_{k>l}{\{\vec{V}_{kl}\rho_{mn}^{\dagger} + h.c\}}\cdot\nabla_{R}\vec{V}_{mn}
\end{equation}
 The time derivative of $\vec{V}_{mn}$ is further approximated by ignoring second derivatives of 
$\vec{V}_{mn}$ on the right hand side of Eq.(21) as well as the terms of the form 
$\vec{V}_{kl}\cdot\nabla_{R}\vec{V}_{mn}$. This amounts to approximating $\dot{\vec{V}}_{mn}$ as
\begin{equation}
  \dot{\vec{V}}_{mn}(R) \approx \nabla_{R}\vec{V}_{mn}(R)\cdot\frac{\vec{P}}{M}
\end{equation}
 The approximation, Eq.(22) can be substituted into Eq.(19) for the second derivative of $R(t)$.
Within this approximation, the sum of contributions from the time derivative of the non-adiabatic
coupling and the corresponding contribution from the derivative of the momentum give rise to a 
magnetic force term. Thus, the following equation is obtained:
\begin{equation}
M\ddot{R}(t) = -\nabla_{R}U(R) -\sum_{m}{\nabla_{R}\epsilon_{m}(R)c_{m}^{\dagger}c_{m}} + 
\sum_{m>n}{\{\vec{V}_{mn}\dot{\rho}_{mn}^{\dagger} + h.c\}} - \frac{\vec{P}}{2M}\times\vec{B} + 
\vec{B}\times\frac{\vec{P}}{2M}  
\end{equation}  
 Here, an effective magnetic field operator has been introduced. It is defined as
\begin{equation}
  \vec{B} = \nabla_{R}\times\sum_{m>k_{f}>n}\{\vec{V}_{mn}(R)\rho^{\dagger}_{mn} + h.c\}
\end{equation}
In the limit where the vibrational coordinate is slowly varying, the derivative of the nonadiabatic
coupling can also be assumed to be small compared to the coupling strength, and hence ignored. If 
this approximation is made, Eq.(23) can be simplified further by substituting Eq.(17) for the derivative 
of the operators $\rho$, and ignoring the magnetic field terms. The substitution gives
\begin{equation}
M\ddot{R} = -\nabla_{R}\tilde{W}(R) + \sum_{k>l}{i\omega_{mn}\{\vec{V}_{mn}(t)\rho_{mn}^{\dagger}(t) 
- \vec{V}_{nm}(t)\rho_{mn}(t)\}} 
\end{equation}                             
 The Born-Oppenheimer potential operator $\tilde{W}$ in Eq.(25) is defined as
\begin{equation}
   \tilde{W}(R) = U(R) + \sum_{k}{\epsilon_{k}(R)c_{k}^{\dagger}c_{k}}
\end{equation}
  Eq.(25) corresponds to a mean-field description of the vibrational dynamics of the molecule. 
The description is a mean-field approximation in the sense that the excitations of electron-hole
pairs are assumed to be on a timescale during which the nonadiabatic vibrational coupling is nearly
constant. This can occur if the vibrational motion has a low amplitude and the nonadiabatic coupling
is otherwise well behaved. The validity of this approximation, and the dynamics in situations where 
this approximation no longer works will be discussed in further detail in a later section.   
\par
  The quantum acceleration in Eq.(25) is the sum of two terms. The first term is the gradient of 
the Born-Oppenheimer potential, and the second term corresponds to off-diagonal hopping terms 
between Born-Oppenheimer potential energy surfaces. The hopping occurs due to the creation or 
destruction of electron-hole pair excitations. If the time dependence of the electron-hole pair 
creation-annihilation operators is explicitly taken into account, a quantum generalized Langevin 
equation (GLE) which describes the reduced dynamics of the vibrational position operator, can be
derived. To do this, Eq.(17) which describes the time dependence of the operators $\rho$ will 
require to be solved. This is solved and a quantum GLE derived below. 
\par
  In what follows, the time dependence of the Born-Oppenheimer excitation energy $\epsilon_{mn}(R)$ 
is neglected. This is mainly for the sake of convenience and this approximation is not crucial to the 
treatment of the problem. The solution to Eq.(17) can be written as 
\begin{equation}
\rho_{mn}^{\dagger}(t) = \rho_{mn}^{\dagger}(0)e^{i\omega_{mn}t} + \frac{1}{M}\int_{0}^{t}{d\tau e^{i\omega_{mn}t}\vec{D}_{nm}(\tau)\cdot\vec{P}(\tau)} 
\end{equation}
  This is substituted into Eq.(25) for the acceleration $\ddot{R}(t)$ to give
\begin{equation}
M\ddot{R}(t) = -\nabla_{R}\tilde{W} - \int_{0}^{t}{d\tau \Gamma(t,\tau)\cdot\vec{P}(\tau)} + \vec{\xi}(t)
\end{equation}
  The "friction" tensor $\Gamma$ is given by
\begin{equation}
 \Gamma^{\mu\nu}(t,\tau) = \sum_{m>n}{\hbar\omega_{mn}\{D^{\mu}_{mn}(t)D^{\nu}_{nm}(\tau)e^{i\omega_{mn}(t-\tau)} + c.c\}}
\end{equation}
  The indices $\mu,\nu$ label the components along the various axes of the coordinate system. 
 The "noise" term $\xi(t)$ is given by
\begin{equation}
\vec{\xi}(t) = \sum_{mn}{\hbar\omega_{mn}\{\vec{D}_{mn}(t)\rho_{mn}^{\dagger}(0)e^{i\omega_{mn}t} - 
\vec{D}_{nm}\rho_{mn}(0)e^{-i\omega_{mn}t}\}}
\end{equation}
  The "noise" operator $\vec{\xi}(t)$ is a purely off-diagonal operator. Hence, its average 
value is zero, i.e,
\begin{equation}
\langle\vec{\xi}(t)\rangle = 0
\end{equation} 
  The position operator of the vibrational mode satisfies a generalized Langevin equation. Under 
certain conditions, this generalized Langevin equation reduces to an equation for quantum Brownian 
motion in the presence of a frictional bath\cite{Gardiner}. The analysis of this equation is 
taken up in the next section.
    
\section*{III. ANALYSIS}
\par  The generalized Langevin equation derived above is an equation relating different quantum 
operators. The effects of the bath influence the system dynamics through the quantum noise 
operator $\vec{\xi}(t)$. In addition, surface electron-hole pair excitations act as a frictional 
drag on the system. Thus, a natural question to ask is whether the fluctuation-dissipation 
theorem is valid for problem if the electron-hole pair excitations are at a thermal equilibrium. 

\par  The relationship between the noise and friction operators can be clarified as below. 
Consider the commutator of the noise operator at different times.
\begin{equation}
[\vec{\xi}(t),\vec{\xi}(\tau)] = \sum_{m>n}\omega_{mn}^{2}\{\vec{V}_{mn}(t)\vec{V}_{nm}(\tau)
e^{i\omega_{mn}(t-\tau)} 
- \vec{V}_{nm}(t)\vec{V}_{mn}(\tau)e^{-i\omega_{mn}(t-\tau)}\}
\end{equation}
  From the definitions, Eq.(29) and Eq.(30) of the noise and friction operators, it is easy to see that 
Eq.(32) implies 
\begin{equation}
  [\vec{\xi}(t),\vec{\xi}(\tau)] = i\hbar\frac{d}{dt}\Gamma(t,\tau) + O(W\dot{W})
\end{equation}
  If, as before, terms of the form $W_{mn}\dot{W_{mn}}$ are neglected, the friction and noise are 
related as
\begin{equation}
  [\vec{\xi}(t),\vec{\xi}(\tau)] \approx  i\hbar\frac{d}{dt}\Gamma(t,\tau) 
\end{equation}
  This relation ensures that the operators $R(t)$ and their conjugate momenta satisfy the Heisenberg
commutation relations\cite{Zoller}. 
  To integrate over the bath degrees of freedom, the assumption made is that the initial density 
matrix of the system is a direct product of the bath density matrix and the system density matrix. 
This means that 
\begin{equation}
  \sigma_{0} \approx \sigma_{s}\otimes\sigma_{e}
\end{equation}
\par This assumption amounts to stating that the system and bath are decoupled at some (initial)
time in the past, and consequently the density matrix of the combined system and bath is factorizable. 
For this problem however, the vibrational and electronic degrees of freedom are never completely 
decoupled. An analogous assumption would be that the vibrational and electronic degrees of freedom 
are adiabatically separable through the Born-Oppenheimer approximation at some time in the past. This 
assumption is consistent with the approximations made earlier in this work. 
  Such an assumption means that the initial electron density matrix $\sigma_{e}$ is a function of the 
vibrational coordinates $R$ as below:
\begin{equation}
 \sigma_{e} = \frac{\exp{(-\beta H({\bf{r}},R))}}{Tr_{r}[\exp{(-\beta H({\bf{r}},R)}]}
\end{equation}
  In terms of an set of Born-Oppenheimer electronic states, the electron density 
matrix is 
\begin{equation}
  \sigma_{e} = \frac{1}{Z(R)}\sum_{k}\mid\phi_{k}\rangle e^{-\beta\epsilon_{k}(R)}\langle\phi_{k}\mid
\end{equation}

   Thus, averages over the electronic degrees of freedom can be defined as traces over the electron 
density matrix $\sigma_{e}$. With this definition, the average of the noise operator over the electron
degrees of freedom is given by
\begin{equation}
  \langle\vec{\xi}(t)\rangle = Z^{-1}(R)Tr\{\sigma_{e}(R)\vec{\xi}(t)\} = 0
\end{equation}
 Furthermore, the adiabatic part of the electron Hamiltonian can be rewritten as 
\begin{equation}
  \sum_{m}{\epsilon_{m}(R)c_{m}^{\dagger}c_{m}} = \sum_{m>n}{\epsilon_{mn}(R)
\rho_{mn}^{\dagger}\rho_{mn}} + \Delta                           
\end{equation}       
   In addition to the constraint $m>n$, the index $m,n$ are such that the corresponding 
energies are further constrained to be above and below the Fermi energy $\epsilon_{f}$, 
i.e $\epsilon_{m} > \epsilon_{f}$ and $\epsilon_{n} < \epsilon_{f}$. The first term on 
the right hand side contains contributions from electronic excitations around the Fermi 
surface, and the second term $\Delta$ consists of the bulk contribution to the energy from 
below the Fermi surface. Since the excitations are of order $1/N$ compared to the total bulk 
electrons, $N$, below the Fermi surface, $\Delta$ can be assumed to be roughly constant. 
With this assumption, the equilibrium distribution of electron-hole pair excitations is 
given by 
\begin{equation}
\langle\rho_{mn}^{\dagger}\rho_{kl}\rangle = \frac{\delta_{mk}\delta_{nl}}
{e^{\beta\epsilon_{mn}} -1}
\end{equation}
   With this in hand, the average of the anticommutator of the noise operator can be written as
\begin{equation}
\langle\{\vec{\xi}(t),\vec{\xi}(\tau)\}\rangle = 2\sum_{m>n}{\hbar^{2}\omega_{mn}^{2}
\{\vec{D}_{mn}(t)\vec{D}_{nm}(\tau)e^{i\omega_{mn}(t-\tau)} + c.c\}(N(\omega_{mn}) + 
\frac{1}{2})}
\end{equation}
  Products of the type $\vec{D}\vec{D}$ are to be understood as tensor products of the 
vectors $D$, i.e the product is a matrix such that its elements are given by, 
$(\vec{D}_{mn}\vec{D}_{nm})_{\mu\nu} = D^{\mu}_{mn}D^{\nu}_{mn}$.  
  The matrix elements $D_{mn}$ are given by the overlap of delocalized electronic wavefunctions 
with their derivatives. The function $N(\omega)$ is the thermal distribution function of the 
electron-hole pair excitations. Within the approximation that they have a Bosonic character, 
it is given by
\begin{equation}
   N(\omega) = [\exp{(\frac{\hbar\omega}{k_{B}T})} - 1]^{-1}
\end{equation}
\subsection*{A. Fluctuation-Dissipation Theorem}
  For the sake of simplicity it is assumed that the product of the off-diagonal matrix elements 
$\vec{V}_{mn}(t)\vec{V}_{nm}(\tau)$ is real in the discussion that follows. With this 
assumption, the electronic friction operator has the form
\begin{equation}
 \Gamma(t,0) = \sum_{m>n}{\hbar\omega_{mn}\vec{D}_{mn}(t)\vec{D}_{nm}(0)\cos[\omega_{mn}t]}
\end{equation}
   The mean-field approximation of the previous section is made that the timescale which governs 
the rate of change of the nonadiabatic coupling matrix elements $D_{mn}(t)$ is considerably larger 
than the timescales governing the creation and annihilation of electron-hole pairs. This assumption 
amounts to asserting that the motion of the molecular degrees of freedom is slow enough to allow for 
the establishment of a local equilibrium distribution for the electron-hole pair excitations. 
Under this assumption, the Laplace transform of the friction operator becomes 
\begin{equation}
  \gamma(s) = \sum_{m>n}{\int_{0}^{\infty}{dt\omega_{mn}\vec{V}_{mn}(t)\vec{V}_{nm}(0)
\cos(\omega_{mn}t)e^{-st}}}
\end{equation}
  If it is assumed that the off-diagonal matrix elements $W_{mn}(t)$ are slowly varying in comparison
to the function $\cos(\omega_{mn}t)$, then the laplace transform becomes
\begin{equation}
\gamma(s) = \sum_{m>n}{\hbar\omega_{mn}\vec{D}_{mn}\vec{D}_{nm}[\frac{1}{s-i\omega_{mn}} + 
\frac{1}{s+i\omega_{mn}}]}
\end{equation}
 This quantity then satisfies the relation
\begin{equation}
 \frac{1}{2}\langle\{\xi(t),\xi(0)\}\rangle = \frac{1}{\pi}\int_{0}^{\infty}{d\omega \Re{[
\gamma(-i\omega + 0^{+})]}\hbar\omega\coth{(\frac{\hbar\omega}{2k_{B}T})}cos(\omega t)}
\end{equation}
   The above relationship is a generalized form of the Callen-Welton fluctuation-dissipation 
theorem\cite{Callen,Hanggi} for quantum systems. It is easy to see that it reduces 
to the classical fluctuation dissipation theorem as $\hbar\rightarrow 0$. 
  If the classical limit for the system variables is taken, the relevant quantity in the 
noise correlation function to be considered is the limiting value 
\begin{equation}
\lim_{\hbar\rightarrow 0} \hbar\omega_{mn}(N(\omega_{mn}) + \frac{1}{2}) = \frac{1}{2}k_{B}T 
\end{equation}
  In this limiting case, the noise correlation function and the electronic friction satisfy the
classical fluctuation-dissipation theorem:
\begin{equation}
   \frac{1}{2}\langle\{\vec{\xi}(t),\vec{\xi}(\tau)\}\rangle = k_{B}T\Gamma(t,\tau)
\end{equation}
  The classical limit here is interpreted to be the condition that the characteristic 
energy of electron-hole pair excitations $\epsilon_{mn}$ is smaller than the temperature
scale, i.e $\epsilon_{mn} \ll k_{B}T$. 
\par  Eq.(46) demonstrates that, given the fairly general assumption of a mean field 
effect of the electronic modes, the resulting dynamics are of a Langevin type for which the 
fluctuation-dissipation relations are satisfied. Unlike traditional Langevin theory, the 
noise and friction are due to a quantum mechanical bath. In the limit where the quantum 
vibrational position and momentum operators are treated classically, the equations of 
motion reduce to a generalization of the Langevin form derived in Ref.(28). This is not 
unexpected because the mean field assumptions made to derive the quantum Langevin equation 
are physically equivalent to a finite temperature Ehrenfest type dynamics in the mixed 
quantum-classical limit.
\subsection*{B. Mean Field Criterion}   
\par Violations of the Callen-Welton fluctuation-dissipation relation are expected to occur when
the timescale which determines the variation of the non-adiabatic coupling is of the
same order of magnitude as the timescale of electron-hole pair excitations. In physical
terms, this means that violations of the fluctuation-dissipation theorem occur 
when the electron-hole pairs created at a given nuclear configuration are unable to relax
into the mean field limit, due to the rapidity of nuclear motion. Quantitatively, this 
criterion amounts to stating that the mean field limit obtains when the change in the 
nonadiabatic coupling over the characteristic timescale for electron-hole pair creation 
is small relative to the value of the coupling. In other words, a mean field Langevin 
description of the noise is retained when
\begin{equation}
\frac{2\pi}{\omega_{mn}}\frac{\mid\dot{\vec{V}}_{mn}\mid}{\mid\vec{V}_{mn}\mid} \ll 1 
\end{equation}
for typical values of the nonadiabatic coupling $\vec{V}_{mn}$ and electron-hole pair 
frequencies $\omega_{mn}$. This criterion can be simplified by rewriting the derivative 
of the coupling in terms of the vibrational velocity. The time period of vibration 
$2\pi/\omega_{mn}$ can be approximated by the time period corresponding to the vibrational 
frequency, $\Omega_{0}$ of the molecule. 
\begin{equation}
\frac{2\pi}{\Omega_{0}}\mid\dot{R}\cdot\nabla_{R}\ln{\mid\vec{V}_{mn}\mid}\mid \ll 1
\end{equation}
   Since this criterion involves time derivatives of the vibrational coordinate, large 
amplitude vibrational motion is more likely to result in violations of the mean-field 
approximation. A more meaningful statement of this criterion can be written in terms 
of the vibrational relaxation rates from the $n=1$ level to the $n=0$ vibrational energy
levels when the local density of states is smooth. 
\par
   Within the Fermi golden rule, the vibrational relaxation rate, $\Gamma$ is characterised
by the proportionality relation;
\begin{equation}  
   \Gamma \propto \sigma^{2}(\epsilon_{f})\langle\mid\vec{V}_{mn}(\vec{R})\mid^{2}\rangle
\end{equation}
   $\sigma(\epsilon_{f})$ is the density of states at the Fermi level. $\langle\mid
\vec{V}_{mn}(\vec{R})\mid^{2}\rangle$ is the transition strength obtained by squaring 
absolute values of electron-hole pair transition matrix elements $\vec{V}_{mn}$ and 
calculating their average weighted with respect to the local density of states for the 
electron-hole pair excitations. 

    If the dependence, on the vibrational coordinate $\vec{R}$, of the local density of 
states $\sigma(\epsilon_{f})$ is ignored, then the mean field criterion for the validity 
of the Langevin description is 
\begin{equation}
   \frac{\pi\nu}{2}\vec{R}_{0}\cdot\nabla_{R}\ln{\Gamma(\vec{R})} \ll 1
\end{equation}
    This inequality depends on the quantum number $\nu$ of the initial vibrational excitation 
of the molecule and the amplitude of the vibrational motion $R_{0}$. This expression is 
derived by replacing the vibrational velocity by the root mean squared velocity of the free 
vibrational motion. The amplitude of the vibration can be written in terms of the vibrational
quantum number to obtain the criterion for the validity of the mean field approximation to be,
\begin{equation}
 \frac{\pi}{2}[\frac{\hbar}{M\Omega_{0}}\nu^{2}(\nu + \frac{1}{2})]^{\frac{1}{2}}
\mid\nabla_{R}\ln \Gamma(\vec{R})\mid \ll 1  
\end{equation}
    It is to be noted that this criterion can be tested experimentally by measuring the 
linewidth near the equilibrium chemisorbate geometry.   

\par The inequality in Eq.(53) is a criterion which provides a quantitative measure of the 
"weak coupling" approximation. When Eq.(53) is satisfied, the adsorbate vibrational 
degrees of freedom obey a generalized Langevin equation subject to quantum white noise.
In the semiclassical limit (i.e when the nuclear degrees of freedom are treated classically) 
the quantum GLE reduces to a standard classical Langevin with an electronic friction.   
\par When the vibrational frequency is sufficiently high to render the mean field approximation
invalid, the dynamics of the molecule gain a strongly nonadiabatic character. In this regime
the approximations wherein the derivatives of the nonadiabatic coupling were ignored are 
invalid. However, the formalism presented here can be extended to account for these corrections. 
Formally, the derivatives of the coupling enter the equations of motion for the acceleration of 
the vibrational coordinate. To derive this, the full contribution to the quantum acceleration is 
retained to give,
\begin{equation}
  M\ddot{\vec{R}} = \dot{\vec{P}} + \sum_{m>\epsilon_{f}>n}\{\vec{V}_{mn}\dot{\rho}^{\dagger}_{mn} 
+ \vec{V}_{nm}\dot{\rho}_{mn}\} + \sum_{m>\epsilon_{f}>n}\{\dot{\vec{V}}_{mn}\rho^{\dagger}_{mn} 
+ h.c\}
\end{equation}
   Unlike in the mean field treatment, the derivatives of the nonadiabatic coupling $V$ no longer 
cancel out of the equation. These derivatives are given by
\begin{equation}
  \dot{\vec{V}}_{kl} = \nabla_{R}\vec{V}_{kl}\cdot\frac{\vec{P}}{M} + \frac{1}{M}
\sum_{m,n}{\{\vec{V}_{mn}\rho^{\dagger}_{mn} + h.c\}}\cdot\nabla_{R}\vec{V}_{kl} -\frac{i\hbar}
{2M}\nabla^{2}\vec{V}_{kl}
\end{equation} 
   The Hamiltonian of the vibrational degrees of freedom in the presence of nonadiabatic coupling 
corresponds to that of a charged particle in a magnetic field. The "vector potential" corresponding 
to this magnetic field is a matrix with matrix elements given by the nonadiabatic coupling 
amplitudes. It is possible to show, with some algebra, that the derivatives of the nonadiabatic 
coupling contribute to this magnetic force.
   
   Thus, in the general case, the resulting equation for the quantum acceleration is 
\begin{equation}
M\ddot{R} = -\nabla_{R}\tilde{W} + \sum_{m>k_{f}>n}i\omega_{mn}\{\vec{V}_{mn}\dot{\rho}^
{\dagger}_{mn} - \vec{V}_{nm}\dot{\rho}_{mn}\} - \frac{1}{2M}\{M\dot{\vec{R}}\times\vec{B} - 
\vec{B}\times M\dot{\vec{R}}\}
\end{equation}
   Eq.(56) must be solved self-consistently with the equations of motion for the electron
pair creation-annihilation operators.  
    
\par The quantum acceleration, Eq.(56), in the strong nonadiabatic coupling regime corresponds 
to a Langevin equation in the presence of a magnetic field. The first two terms correspond to the 
mean field contributions to the quantum acceleration. In the mean field limit, the two Lorentz 
force terms vanish. In this equation, the magnetic field is itself a matrix operator which is 
a linear combination of electron-hole pair creation and annihilation operators. Physically, 
this implies that each electron-hole pair excitation is associated with a corresponding magnetic 
flux. This magnetic flux is generated through the nonadiabatic coupling vector associated with 
the electron-hole pair. Hence, the motion of the molecular coordinates can be thought of as 
the motion of a charged particle moving in a fluid (electron-hole pair excitations). As the 
particle velocity increases, it creates eddy currents due to the magnetic coupling which aid 
in dissipating the energy of the particle. A clearer interpretation of this is as follows. 
For sufficiently weak nonadiabatic coupling, the vibrational motion of the molecule causes
electrons from near the Fermi energy to be excited and form electron-hole pairs. As the 
nonadiabatic coupling grows in strength, multiple electron-hole pairs as well as electrons 
with energies that are farther below the Fermi energy begin to be excited. This is manifested 
through the magnetic field contribution to the acceleration. For very strong nonadiabatic coupling, 
the excitations could involve sufficient energy to allow for an electron to escape the metal 
surface.    
 
\par Eq.(56) corresponds to a set of nonlinear differential equations for the quantum 
acceleration. This nonlinearity is due to the coupling of the forces and accelerations 
corresponding to different directions of the vibrational coordinate. There is also a further 
source of nonlinearity in that this equation is coupled to the equations of motion for the 
electron-hole pair operators. Consequently, a general solution of this set of equations is 
difficult. A quantitative description of the molecular dynamics in the presence of 
strong nonadiabatic coupling is currently under investigation.  
      
\section*{IV. DISCUSSION}
    In this work, rigorous quantum equations of motion have been derived to describe
the quantum dynamics of molecules on metal surfaces when subject to a coupling to 
surface electron-hole pair excitations. It is found that as long as the nonadiabatic 
coupling has a slow time variation relative to the timescale of non-adiabatic 
excitations, the reduced quantum dynamics of the adsorbed molecules are described by a 
quantum version of the generalized Langevin equation. Further, within this approximation
it has been shown that the quantum noise and friction operators satisfy the generalized
version of the Callen-Welton fluctuation dissipation theorem. In the semiclassical limit
wherein the molecular modes are treated classically while retaining the quantum 
mechanical nature of the electron-hole pair bath, the noise and friction operators 
satisfy the classical fluctuation-dissipation relation. 
\newline
\par
   The classical limit of the Langevin equation is found to obtain if the characteristic 
energy of electron-hole pair excitations is much smaller than the ambient temperature of 
the surface. However, at low temperatures, this no longer holds and the dynamics should be 
described correctly by the quantum Langevin equation. Since dissipative effects due to 
electron-hole pair excitations are dominant over other mechanisms only at low temperatures, 
this implies that quantum effects could contribute significantly to molecular motion on the 
surface. The classical limit also enables a physical interpretation of the Langevin equation. 
The electronic energy levels of the adsorbed molecule broaden due to interaction with the 
surface. Due to this, a part of the broadened resonance goes below the Fermi energy and becomes 
populated with metallic electrons. Molecular vibration causes a dynamical charge transfer 
because the degree of overlap of the resonance below the Fermi level changes\cite{Persson2}. 
If the electrons lag the nuclear motion, the resonance can be unoccupied at or below the 
fermi level, thus causing an electron-hole pair excitation and resultant dissipation. The
connection between the Langevin formulation and the dynamical charge transfer become 
explicitly clear in the Markovian limit, where the electronic friction kernel becomes 
proportional to the linewidth of the vibrational excitation given by the Fermi golden rule.     
\newline
\par 
   It should be noted that numerical computations of the Langevin equation, even in the 
classical limit are a challenging task. In addition to the difficulties associated with 
the simulation of Brownian processes with memory dependent friction, additional problems 
arise due to the dependence of the friction kernel on the vibrational coordinates. A full
solution would necessarily involve a self consistent propagation of the Langevin equation 
with the vibrational motion. However, the numerical simulation of the equation becomes more 
tractable in the Markovian limit, where the memory dependence of the friction kernel is 
dropped. In this situation, approximate values of the friction term can be obtained through 
the use of density functional theory or other electronic structure methods to estimate the 
matrix elements of the coupling potential around the Fermi level and thus evaluating the 
electronic friction. To obtain full self consistency with vibrational motion, it is necessary 
to perform this calculation every few time steps. Thus, full self consistency requires 
considerable computational expense and limits the time for which the vibrational dynamics 
can be propagated.
\newline 
\par
   A simplification of the generalized Langevin equation, Eq.(28) obtains when the local electron
density of states is smooth and slowly varying. In this circumstance, the surface electronic 
structure provides a constant frictional background and the memory dependence of the friction
kernel can be dropped. Thus, under such conditions, the Markovian limit of the GLE is obtained. 
However, the presence of sharp features in the local density of states causes the dynamics to 
become non-Markovian and the friction kernel becomes memory dependent. This change is because 
the friction kernel is a two point sum over non-adiabatic couplings evaluated at vibrational 
positions corresponding to different times. If the electron density of states is heterogenous 
and sharp around one of the two vibrational positions, the summation over local electronic levels 
are different for the two times, and the friction kernel becomes memory dependent. This memory
dependence is in general difficult to treat, however if the sharp features in the density 
of states are isolated peaks in a smooth background, approximations could be developed for the 
numerical resolution of these difficulties by treating the peaks as perturbations in the 
Fourier representation of the classical limit Langevin equation. These and other issues concerning 
the computational study of the dynamics within the framework presented here are presently 
under consideration and will be discussed in more detail at a later date     
\newline         
\par
   In the regime of strong nonadiabatic coupling, the equations of motion for the 
vibrational coordinate become nonlinear, and correspond to the dynamics of a charged particle 
in the presence of both electric and magnetic field environments. Contrary to the mean field 
regime, a strong nonadiabatic coupling is likely to enable multiple electron-hole pair excitations
as well as electron-hole excitations with high energy. This suggests that the dissipation of energy 
in the presence of strong nonadiabatic coupling is likely to be nonexponential, and the corresponding 
diffusion of the vibrational coordinate is no longer a purely Brownian process. This is consistent 
with the evidence provided by a recent series of experiments\cite{Wodtke1}$^{-}$\cite{Wodtke4}. In 
these experiments, highly vibrationally excited $NO$ molecules were scattered off a metal surface 
and it was observed that the scattering process resulted in the relaxation of the vibrational state 
by several vibrational quantum numbers. In one of these experiments, the scattering process was 
found to result in the emission of electrons from the surface\cite{Wodtke3}. It is therefore 
suggested that the equations of motion developed here could be used to quantitatively simulate 
the dynamics of high vibrationally molecules on the metal surface and test these possible 
dissipation mechanisms.
\newline
\par
   The formalism presented here is quite general and the equations of motion derived here are 
rigorous and based on first principles. Furthermore, this method can be easily modified to 
study nonadiabatic molecular dynamics in more general contexts. When the nonadiabatic dynamics 
occurs with coupling between discrete sets of Born-Oppenheimer states, Eq.(56) remains the same. 
However, the electron pair creation operators can no longer be approximated to be bosonic, and 
will have to be propagated using the exact equations of motion. Since these operators are closely 
linked to the electron density matrix, the equations of motion can be approximated with Pauli type 
master equations. Under such circumstances, it is intuitively apparent that existing mixed-quantum 
classical recipes correspond to different approximate solutions of these master equations. As a 
natural consequence, when the number of electrons are taken to infinity, different mixed quantum 
classical methods are likely to converge to the mean field GLE derived here. The methodology 
presented here can therefore aid in the further development of more refined mixed quantum classical 
molecular dynamics methods. The application of this methodology to study mixed quantum-classical 
dynamics will be presented in a future publication.  

    
\section*{V. Acknowledgements}
    Financial support from a Yale University Graduate Teaching fellowship is gratefully 
acknowledged. I thank Prof John Tully for teaching me much of this subject and for his 
generous encouragement and support. I also am grateful to Dr A. V. Madhav for useful 
comments and discussions.  


\end{document}